# Assessment of scattered and leakage radiation from ultra-portable digital chest X-ray systems: An independent study

Leonie E Paulis[1,2,3,*], Roald S Schnerr[1,4], Jarred Halton[1], Zhi Zhen Qin[2], Arlene Chua[5]

[1] Médecins Sans Frontières, International, Amsterdam, the Netherlands

[2] Department of Digital Health, Stop TB, Geneva, Switzerland

[3] Department of Medical Physics, Maxima Medical Center, Veldhoven, the Netherlands

[4] Department of Radiology and Nuclear Medicine, Maastricht University Medical Center, Maastricht, the Netherlands

[5] Médecins Sans Frontières, International, Geneva, Switzerland

* Corresponding author

E-mail: leonie.paulis@amsterdam.msf.org (LP)

# Abstract

Ultraportable X-ray devices are ideal for TB screening in resource limited settings. Unfortunately, guidelines on the radiation safety of these devices are lacking. The aim of this study was to determine the radiation dose by scattered and leakage radiation of four ultraportable X-ray devices to provide a basis for these guidelines.

Radiation dose measurements were performed with four ultraportable X-ray devices that meet the WHO/IAEA criteria. An anthropomorphic thorax phantom was positioned in posterior-anterior orientation in a clinically representative X-ray set-up. X-ray exams were acquired with the following scan parameters: 90 kV, 2.5 mAs and maximum mAs, 1 m and 1.8 m source-skin-distance. The entrance skin dose was measured at the center of the phantom. The scattered radiation dose was measured at 1 m from the phantom as a function of scatter angle. Leakage radiation was measured at 0.5 m from the X-ray tube with collimators closed and covered with additional lead.

From the scatter measurements the ‘safe distance bunny’ was determined, which was the minimum distance to the phantom to stay below the international dose limit to the public (1 mSv/year) at a given workload: longest distance (related to highest scatter dose) was observed behind the edges of the detector and back towards the X-ray tube, whereas shortest distance (related to lowest scatter dose) was observed to the sides of the phantom.

For the radiographer position, the total radiation dose by scattered and leakage radiation was determined in various scenarios. In most cases, the total radiation dose of ultraportable X-ray devices can be kept below 1 mSv/year by employing basic radiation safety rules: 1. reduce time in the X-ray field, 2. increase distance to the X-ray source, and 3. use shielding measures (e.g. lead apron).

Ultraportable X-ray devices can be safely used for TB screening when using adequate precautions.

# Introduction

Tuberculosis (TB) remains a significant global health crisis, surpassing all other infectious diseases in terms of annual mortality, despite being treatable [1]. To combat TB, early diagnosis is critical, e.g. by systematically screening for TB in subpopulations of people who are at higher risk of being exposed to TB or with structural risk factors for TB [2,3]. In low- and middle-income countries (LMICs), bearing the brunt of the TB burden, constrained healthcare systems often resort to symptom-based screening, despite its low sensitivity, due to the high cost of more sensitive screening tools. At least half of the TB patients would have been missed if only screened with symptoms alone, as evidenced by recent prevalence surveys [3-5]. Therefore, there is a need for more sensitive tools that facilitate large-scale TB screening in resource-limited conditions [3].

Chest X-ray (CXR) imaging is a highly sensitive diagnostic technique, that is capable of detecting even asymptomatic TB cases [3]. However, traditionally CXR imaging required a dedicated radiology infrastructure, thereby limiting its use in communities. Recently, advancements in radiological equipment have led to the development of ultraportable (UP) X-ray devices, which are designed to fit within a suitcase or backpack and be moved regularly to areas of need [6,7]. These systems are composed of a low-weight, battery-powered X-ray generator and a highly sensitive and dose-efficient digital detector. They can be paired with artificial intelligence powered computer-aided detection (CAD) software that provides automated and standardized interpretation of CXR without human interpretation [8-12]. This unique set-up makes UP X-ray devices with CAD software an ideal tool to facilitate TB screening in places where even conventional infrastructure (e.g. stable power supply) is non-existent [1,13].

Recent studies have shed light on the challenges faced when using UP X-ray devices for TB screening, with a main concern being the absence of radiation safety guidelines [9]. Existing international regulatory standards are primarily designed for traditional X-ray machines in a dedicated hospital area, e.g. a permanent radiology department [14]. This restricts the potential of UP X-ray systems to be used in conditions without adequate healthcare infrastructure, which is their primary intended use [6]. Therefore, it is important to develop guidelines for safe use of UP X-ray devices, to minimize radiation related health risks.

From a radiation safety perspective, the health risk to the individual patient by CXR for TB screening is negligible due to the low X-ray radiation dose needed (equivalent to a few days of natural background radiation) [15]. Particular focus should be on the radiation safety of radiology staff and others that are regularly in the close vicinity of UP X-ray devices during their operation. Although they are not exposed to the primary X-ray beam (which is focused on the patient), they are exposed to scattered and leakage radiation. Scattered radiation is X-ray radiation that, upon interaction with the patient, is scattered from within the patient to its surroundings, while leakage radiation emanates from the X-ray tube's protective housing [16]. Both scattered and leakage radiation are much lower in intensity than the primary X-ray beam [16], however, the high patient throughput in a TB screening setting may result in a substantial cumulative radiation dose to people working there as opposed to people that spend only limited time at the TB screening site (e.g. people in waiting rooms). To minimize its health risks, radiation exposure levels are subjected to international radiation dose limits of an effective dose of 1 mSv/year to the public and up to 20 mSv/year for radiology staff (S1 and S2 Tables) [14,17]. International guidelines require individual dose monitoring when radiology staff is expected to receive more than 1 mSv/year [17].

To gain an insight in the radiation safety of UP X-ray devices for TB screening, the current study conducted an independent analysis of both the scattered and leakage radiation from CXR exams by all commercially available UP X-ray systems that met the criteria from the World Health Organization (WHO) and the International Atomic Energy Agency (IAEA) at the time of the study [6]. The objective of this study is to contribute to guidance on the safe use of UP X-ray devices for TB screening in areas outside dedicated X-ray departments, while complying with internationally recognized standards [14,17].

# Methods

## Experimental set up and radiation dose measurements

Four UP X-ray devices were selected that met the WHO/IAEA criteria at the time of the study. Device characteristics are shown in Table 1.

**Table 1. UP X-ray device characteristics.**

| System* | Range | | Recommended settings for CXR | | |
|---|---|---|---|---|---|
| | **kV** | **mAs** | **kV** | **mAs** | **SSD [m]** |
| Fuji | 50-90 | 0.20-2.50 | 90 | 0.5 | 0.8 |
| MinXray | 40-90 | 0.2-20.0 | 90 | 1.0 | 1.6 |
| Sinopharm | 40-100 | 0.4-50.0 | 90 | 2.5 | 1.3 |
| Delft Imaging | 40-90 | 0.1-20.0 | 90 | 1.2 | 1.3 |

kV = tube voltage, mAs = tube current * time product, and SSD = source skin distance, which is source image distance - phantom thickness (20 cm). Parameters are explained in S3 Table.

* Fuji: FDR Xair (Fuji Film, Japan), MinXray: Impact Wireless (MinXray, USA), Sinopharm: SR-1000 (Shantou Institute of Ultrasonic Instruments Co. Ltd. and Sinopharm Biotech, China), Delft Imaging: Delft Light (Delft Imaging, Netherlands)

The study was conducted in the Department of Radiology and Nuclear Medicine, Maastricht University Medical Center (the Netherlands). The UP X-ray devices were positioned in a clinically representative setup for CXR with an anthropomorphic thorax phantom (Alderson phantom (Radiology Support Devices Inc.,USA)) positioned at chest height (1.4 m) in posterior-anterior (PA) orientation with respect to the X-ray tube and the X-ray detector positioned at its anterior side (Fig 1). The X-ray beam was collimated to the phantom's thorax. Basic quality control of the primary X-ray beam was performed prior to the measurements.

**Fig 1. Schematic top view of the experimental set-up.** ESD measurements were performed at the posterior side (in front) of the phantom (black dot). Scattered radiation measurements were performed at 1 m from the center of the phantom at 30º intervals (dark grey dots). Leakage radiation measurements were performed at 0.5 m from the X-ray tube to the left, right and back (light grey dots). The set-up for SSD 1.8 m is shown (for SSD 1 m the tube position coincided with the 180º position).

Radiation dose measurements were performed as described below using a Piranha Multi dosimeter version 5.7 (RTI group, Sweden) with the external dose probe connected (air kerma dose range 0.1 nGy – 1.5 kGy, air kerma accuracy 5%). When applicable, the radiation dose was converted from Gy to Sv using a conversion factor of 1.4 (S1 Table) [18]. All results were anonymized with respect to the individual X-ray systems.

## Entrance skin dose (ESD)

To determine the X-ray radiation output of the UP X-ray devices for a CXR exam, the ESD was measured with the dosimeter on the posterior side of the phantom in the center of the field of view (where the X-ray beam entered the phantom) (Fig 1; black dot). ESD was measured at 90 kV for a range of mAs values (0.5 mAs to maximum mAs of each individual system) and SSD of 1 m and 1.8 m.

## Scattered radiation dose

To characterize the scattered radiation dose as a function of scatter angle, dose measurements were performed with the dosimeter at chest height at 1 m from the center of the phantom at 30º increments (Fig 1; dark grey dots). CXR exams were made with scan parameters 90 kV and both 2.5 mAs and maximum mAs, for an SSD of 1 m and 1.8 m. The setting of 2.5 mAs was selected as an indicative mAs at the upper limit of the normal range of mAs values recommended by manufacturers for clinical use, whereas maximum mAs was selected to illustrate the worst case scenario.

## Leakage radiation dose

Leakage radiation dose was measured according to IEC 60601: behind and to the left and right of the X-ray tube (Fig 1; light grey dots). The collimator was closed and blocked with 5 cm thick lead blocks. Measurements were performed with the dosimeter at 0.5 m from the X-ray anode at 90 kV and maximum mAs for the system, with a limit of 10 mAs. The leakage dose at 0.5 m was converted to the leakage dose at 1 m using the inverse square law:

$$\frac{D_{L,0.5m}}{D_{L,1m}} = \frac{(1m)^2}{(0.5m)^2},$$

with $D_{L,0.5m}$ = measured leakage dose per exam at 0.5 m from the X-ray tube [Gy] and $D_{L,1m}$ = calculated leakage dose per exam at 1 m from the X-ray tube [Gy].

Subsequently, per system, the maximum leakage dose of the three measurements was compared to international dose limit for leakage radiation (1 mGy/h).

# Calculation of yearly radiation dose

## Scattered radiation dose

The yearly scatter dose at 1 m from the phantom was calculated from the scatter dose measurements of a single CXR exam at 90 kV and 2.5 mAs, for both SSD 1 m and 1.8 m. This was done for a workload of 50, 100 and 200 exams per day, by multiplying, for each angle, the median value of the four systems with the total number of exams per year (assuming 5 days/week, 52 weeks/year):

$$D_{S_year,1m} = 1.4 \cdot \breve{D}_{S,1m} \cdot W \cdot 5 \cdot 52 \,,$$

with $D_{S_year,1m}$ = calculated yearly scatter dose at 1 m from the phantom [Sv], $\breve{D}_{S,1m}$ = median measured scatter dose per exam at 1 m from the phantom [Gy], W = workload [exams/day], 1.4 = conversion factor Gy to Sv [18].

Subsequently, for each angle, the distance from the center of the phantom was calculated at which the yearly scatter dose is 1 mSv/year ($r_{1mSv}$), which is the international public radiation dose limit [14,17]. This was done using the inverse square law:

$$\frac{D_{S_year,1m}}{1mSv} = \frac{{r_{1mSv}}^2}{(1m)^2} \,,$$

From this, 1 mSv/year isodose lines were plotted that showed for each angle the minimum distance to the phantom to stay below the public radiation dose limit.

## Leakage radiation dose

The yearly leakage dose at 1 m from the X-ray tube was calculated for a workload of 50, 100 and 200 exams per day, by multiplying the median value of the maximum dose per system with the total number of exams per year (assuming 5 days/week, 52 weeks/year):

$$D_{L_year,1m} = 1.4 \cdot \breve{D}_{L,1m} \cdot W \cdot 5 \cdot 52 \,,$$

with $D_{L_year,1m}$ = calculated yearly leakage dose at 1 m from the X-ray tube [Sv], $\breve{D}_{L,1m}$ = median calculated leakage dose per exam at 1 m from the X-ray tube [Gy], W = workload [exams/day], 1.4 = conversion factor Gy to Sv.

## Total radiation dose at typical radiographer position

The total yearly radiation dose by scattered and leakage radiation at a typical radiographer position was determined at a distance of 1 m and 2 m behind the X-ray tube for 90 kV and 1 mAs, 2.5 mAs and 10 mAs, for both SSD 1 m and 1.8 m.

First, the scatter and leakage doses per exam were calculated at 1 m and 2 m from the X-ray tube. For scattered radiation, the median scatter dose per exam measured at 1 m from the phantom in the direction of the X-ray tube (180°) was used for 2.5 and 10 mAs. For 1 mAs (which was not measured), linear extrapolation was used to obtain the dose per exam. These doses were converted to the dose at 1 m or 2 m from the X-ray tube using inverse square law:

$$\frac{\breve{D}_{S,1m}}{D_{S,r_{tube}}} = \frac{(r_{phantom_staff})^2}{(1m)^2},$$

with $\check{D}_{S,1m}$ as defined above, $D_{S,r_tube}$ = calculated scatter dose per exam at distance '$r_{tube}$' from the X-ray tube [Gy], $r_{phantom_staff}$ = SSD + $r_{tube}$ [m], with $r_{tube}$ = 1 m or 2 m.

For leakage radiation, the median value of the maximum leakage dose per system at 1 m from the X-ray tube was used for 10 mAs. For 1 and 2.5 mAs (which were not measured), linear extrapolation was used to obtain the dose per exam. To calculate the dose at 2 m from the X-ray tube, the dose at 1 m was converted using inverse square law:

$$\frac{\breve{D}_{L,1m}}{D_{L,2m}} = \frac{(2m)^2}{(1m)^2},$$

with $\check{D}_{L,1m}$ as defined above, and $D_{L,2m}$ = calculated leakage dose per exam at 2 m from the X-ray tube [Gy].

Additionally, to illustrate the effect of wearing a 0.25 mm single layer lead equivalent protective apron and thyroid collar, both the scatter and leakage dose per exam were converted with a correction factor of 5 [19].

Subsequently, the yearly scatter and leakage dose were calculated from the scatter and leakage dose per exam for a range of workloads (0 to > 1000 exams per day) as described previously (assuming 5 days/week, 52 weeks/year). The total dose at a typical radiographer position (1 m and 2 m behind the X-ray tube) was defined as the sum of the yearly scatter and leakage doses.

$D_{radiographer_year} = D_{S_year} + D_{L_year}$,

From these data, the number of exams per day was determined at which the total dose exceeded the international public dose limit (1 mSv/year).

# Results

## Entrance skin dose

The ESD of a single PA CXR exam is shown in Fig 2. Fig 2A shows the ESD of the individual UP X-ray systems for 0.5-10 mAs (SSD 1.8 m, 90 kV). Within this mAs-range, the ESD of all systems was lower than or equal to the reported ESD from conventional X-ray systems (0.006-0.15 mGy vs. 0.15 mGy) [20]. The ESD increased linearly with increasing mAs ($r^2$=1 for each system), which is in accordance with literature [21]. The effect of SSD on ESD is shown in Fig 2B. The ESD decreased with increasing SSD, following the inverse square law [21], with a deviation of max 10%, caused by inherent inaccuracy of the dosimeter (5%) and experimental set-up.

**Fig 2. ESD of a single PA CXR phantom exam.** A. ESD at 90 kV and SSD 1.8 m for all systems. B. ESD of system 1 (with median ESD of all systems) at 90 kV for SSD 1 m and 1.8 m.

## Scattered radiation dose

The scattered radiation dose from a single PA CXR exam at 1 m from the Alderson phantom is shown in Fig 3 (90 kV, 2.5 mAs). Differences in scatter dose between UP X-ray systems were relatively small. For all systems, highest scatter dose was observed in the direction back towards the X-ray tube (180°; *i.e.* backscatter) and behind the detector's edges (30° and 330°), both positions where X-rays encounter little attenuation by tissue before they are reflected out of the phantom. At these positions the scatter dose ranged from 0.2-0.6 µSv for SSD 1.8 m and 0.6-2.0 µSv, for SSD 1 m. The lowest scatter dose was observed behind the center of the detector (0$^{o}$) and perpendicular to the X-ray primary beam (270-300$^{o}$ and 60-90$^{o}$), because of higher attenuation by the larger tissue mass that has to be traversed and the inherent physics of scatter as described by the Klein-Nishina formula [22]. At these positions the scatter dose ranged from 0.0-0.2 µSv for SSD 1.8 m and 0.1-0.7 µSv, for SSD 1 m. The scatter dose was lower for SSD 1.8 m compared to SSD 1 m (Fig 3A & 3C vs. Fig 3B & 3D) and increased linearly with increasing mAs from 2.5 mAs to 10 mAs.

**Fig 3. Scatter dose per exam at 1 m from Alderson phantom for all systems.** A & C. Scatter dose at 90 kV, 2.5 mAs and SSD 1.8 m. B & D. Scatter dose at 90 kV, 2.5 mAs and SSD 1 m. In C & D the arrows indicate the angular distribution, which is identical to Fig 1, with each point representing a 30$^{0}$ increment counter clockwise from 0$^{o}$ to 360$^{o}$ (upper y-axis).

In Fig 4 the 1 mSv/year isodose lines are shown that correspond to the distance from the Alderson phantom for each angular position (0$^{o}$-360$^{o}$) at which the yearly scatter dose exceeded the international public radiation dose limit (at 90 kV, 2.5 mAs) [14,17]. The isodose lines enclose a 'bunny' shape around the phantom: The longest distance (corresponding to the highest scatter dose measured at 1 m) was observed in the direction back towards the X-ray tube (180$^{o}$, the bunny's chin) and behind the edges of the detector (30$^{o}$ and 330$^{o}$, the bunny's ears). The shortest distance (corresponding to the lowest scatter dose measured at 1 m) was observed behind the center of the detector (0$^{o}$) and to the left and right of the phantom (60-90$^{o}$ and 270-300$^{o}$, the bunny's cheeks).

**Fig 4. 1 mSv/year isodose 'bunnies' for various workloads.** A & B. Minimum distance to the phantom for 90 kV, 2.5 mAs and SSD 1.8 m and 1 m, resp. at which the yearly scattered radiation dose was below 1 mSv/year. P and X indicate position of the phantom and X-ray tube. The angular distribution is identical to Fig 1.

At the position with highest scatter dose (*i.e.* back towards the X-ray tube (180$^{o}$)), a workload of 50 exams/day resulted in a 1 mSv/year-distance from the center of the Alderson phantom of 2.6 m for SSD 1.8 m and 4.5 m for SSD 1 m. Increasing the workload to 200 exams/day, increased the 1 mSv/year-distance to 5.1 m for SSD 1.8 m and 8.9 m for SSD 1 m. At the position with lowest scatter dose (*i.e.* to the left and right of the phantom (60-90$^o$ and 270-300$^o$), a workload of 50 exams/day resulted in a 1 mSv/year-distance from the center of the Alderson phantom of 1.5 m for SSD 1.8 m and 2.5 m for SSD 1 m. Increasing the workload to 200 exams/day, increased the 1 mSv/year-distance to 3.0 m for SSD 1.8 m and 5.1 m for SSD 1 m.

## Leakage radiation dose

For all systems, the leakage radiation dose rate of a single CXR exam was below the international limit of 1 mGy/h. The leakage dose of a single CXR exam at 1 m from the X-ray tube was only minor (0-0.1 µGy) (Table 2). However, at high daily workloads, the yearly cumulative leakage dose at 1 m was substantial, ranging from 0.6 to 2.4 mSv/year for 50 to 200 exams/day. Leakage radiation is independent from SSD.

**Table 2. Leakage of primary radiation from the X-ray tube (90 kV, 2.5 mAs, 1 m from X-ray tube).**

| | **Leakage radiation [mGy]** | | |
|---|---|---|---|
| | **Position vs. X-ray tube** | | |
| | **Left** | **Right** | **Behind** |
| **System 1** | 1.3E-05 | 3.0E-05 | 3.1E-05 |
| **System 2** | 3.8E-05 | 1.2E-04 | 0* |
| **System 3** | 1.4E-05 | 1.1E-05 | 0* |
| **System 4** | 2.2E-05 | 3.4E-05 | 2.5E-05 |

* Below the detection limit of the dosimeter.

## Total radiation dose at typical radiographer position

The yearly total radiation dose by scattered and leakage radiation was determined at an expected position of the radiographer, which would typically be 1-2 m behind the X-ray tube. In Fig 5 the maximum number of exams is shown that can be performed while keeping the yearly total dose at this position below the international public radiation dose limit (1 mSv/year). The effect of various parameters is illustrated: Increasing SSD increased the maximum number of exams (Fig 5A vs. Fig 5B), e.g. at 1 m from the X-ray tube (at 90 kV, 2.5 mAs, no lead apron), the number of exams increased from 9 to 35, when the SSD was increased from 1 m to 1.8 m. Increasing mAs decreased the maximum number of exams. For example, at 1 m from the X-ray tube (90 kV, SSD 1.8 m, no lead apron), the number of exams decreased from 88 to 9 exams, when mAs was increased from 1 mAs to 10 mAs. Increasing the distance to the X-ray tube from 1 m to 2 m increased the maximum number of exams, e.g. the number of exams increased from 35 to 83, when the distance increased from 1 m to 2 m (90 kV, 2.5 mAs, SSD 1.8 m, no lead apron). Wearing a lead apron increased the maximum number of exams before exceeding the 1 mSv-limit. In the previous example (90 kV, 2.5 mAs, SSD 1.8 m), wearing a lead apron increased the maximum from 35 exams to 175 exams at 1 m from the X-ray tube and from 83 exams to 413 exams at 2 m from the X-ray tube.

**Fig 5. Number of exams that can be performed per day with the total radiation dose by scattered and leakage radiation at the typical position of the radiographer < 1 mSv/year.** A. Scan parameters were SSD 1.8 m and 90 kV. B. Scan parameters were SSD 1 m and 90 kV. The white bars illustrate the effect of wearing a lead apron and thyroid collar, using a correction factor (5x) as published in [19].

# Discussion

UP X-ray devices are an ideal tool to facilitate TB screening in settings without adequate healthcare infrastructure, but radiation related health risks to the radiographers and others near the X-ray set-up must comply with international safety standards [14,17]. Our results show that for most realistic TB screening scenarios, it is possible to keep the total radiation dose by scattered and leakage radiation below the international limit to the public (1 mSv/year) if proper precautions are taken (Fig 5).

Our study was conducted in an experimental environment. Real life conditions may have variable set ups in terms of X-ray tube and detector alignment, distance, collimation and scan parameters, that may influence the results.

In clinical practice, the radiation dose to the patient and staff should be as low as reasonably achievable (ALARA) without comprising diagnostic image quality [23]. This is achieved by optimizing X-ray scan parameters (*i.e.* kV, mAs, SSD) for each individual patient (S3 Table). For UP CXR systems, the WHO/IAEA recommend a tube voltage of at least 90 kV to obtain sufficient image contrast [6]. For a given kV, the appropriate mAs depends on patient size (*i.e.* increasing mAs with increasing size). Furthermore, mAs is related to PA CXR distance, which is typically between 1-1.8 m, with a longer distance requiring a higher mAs. For UP X-ray devices, this requires careful manual manipulation of scan parameters by radiographers since these devices are not equipped with automatic exposure control.

To keep the radiographer's radiation dose as low as possible, additional radiation protection measures are strongly advised. A possible way to reduce the radiation exposure would be to reduce the time spent in the radiation field, e.g., by alternating work shifts or rotating roles in the TB screening program [14]. Furthermore, the distance to the radiation source *i.e.* patient for scattered radiation and X-ray tube for leakage radiation should be maximized (2x distance = $1/(2^2)$x dose). For this purpose, the scattered radiation pattern (the 'bunny') can aid in the design of CXR screening sites to facilitate a set-up that fits local conditions: Areas to the patients' left and right are ideal for patient administration or a waiting area, whereas the area behind the edges of the detector should be avoided, especially since at this location the radiation dose could be significantly higher if the primary X-ray beam is not consistently collimated (focused) on the patient. Importantly, adequate means should be provided to the radiographers to ensure sufficient distance can be created e.g. a long cord for the X-ray exposure switch. From this perspective, using handheld X-ray systems as a true handheld (*i.e.* not mounted to a tripod) at the workload that is expected in TB screening programs will be suboptimal in terms of radiation safety. Finally, the IAEA advises the use of protective clothing (*i.e.* lead aprons and thyroid collars) when X-ray exams are performed outside of a dedicated X-ray facility [14], as is the intended use of UP devices [6]. Our results indicated that for most clinical scenarios lightweight lead gowns of 0.25 mm lead equivalent material provide sufficient radiation protection. Heavier lead gowns (e.g. 0.5 mm lead equivalent material) may not necessarily be required in these settings. Moreover, a potential disadvantage of heavier lead gowns, especially when used in hot and humid climates, maybe a decreased wearer compliance.

When staff members are at risk of a radiation exposure above 1 mSv/year, it is strongly advised to monitor staff with personal radiation dosimeters, so that prompt action can be taken when required [17,22]. In some LMICs, access to reliable personal dosimetry services may be limited [13]. Fortunately, our results indicate that in most UP X-ray settings, the radiation dose can be kept below 1 mSv/year.

An important implication of this study is the necessity to ensure comprehensive training of local TB screening staff in not only radiation safety principles, but also CXR techniques, including the ALARA principle and collimation of the X-ray beam to the thorax [13]. Normal clinical use according to ALARA implies adjusting X-ray scan parameters, especially mAs, because too high mAs poses an unnecessary dose on both patients and radiographers without an additional diagnostic benefit. Training is crucial considering individuals with variable radiography expertise are involved in CXR screening programs, and there is a limited support network in LMICs [24].

# Conclusion

Our study analyzed the radiation doses from UP CXR exams, which shows that UP X-ray devices can be safely used for community TB screening in resource-limited settings following proper guidance. Depending on a site's expected patient throughput, additional radiation safety measures such as reducing exposure time, increasing distance to the radiation source, or wearing protective clothing should be considered in initial site design to avoid an unnecessary health risk for radiographers or those regularly within the vicinity of UP X-ray devices during their operation.

# Acknowledgements

We would like to thank all manufacturers for loaning their devices to our independent testing of their systems to obtain data under identical conditions. The manufacturers had no role in the study design, data collection, analysis and interpretation. Additionally, we are grateful to Maastro and Maastricht UMC+ for providing the equipment and facilities to perform our study.

# Supporting information

**S1 Table. Description of radiation dose units [23].**

**S2 Table. Radiation dose limits [23].**

**S3 Table. Description of X-ray scan parameters [21].**

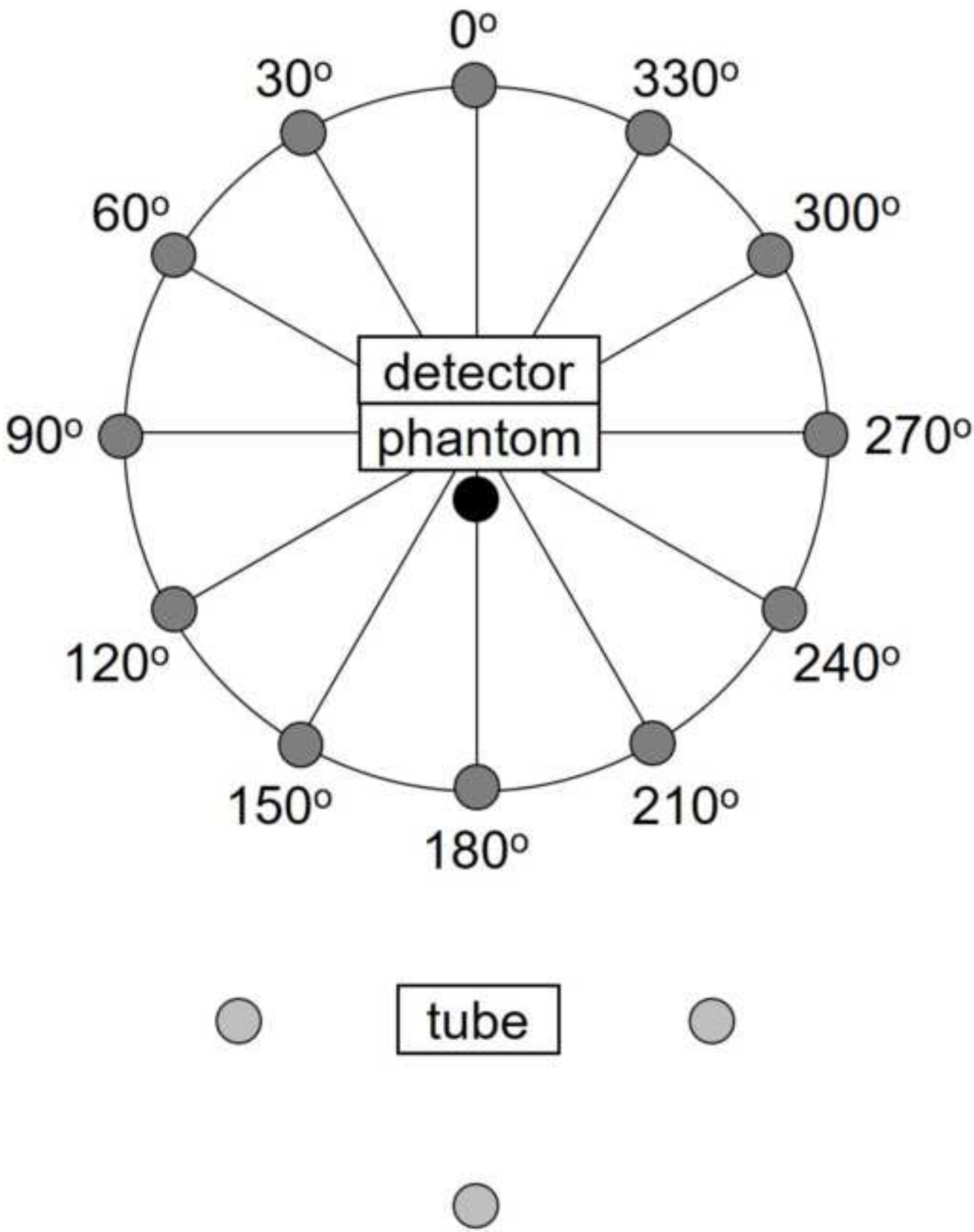
0°
30°
330°
60°
300°
detector
phantom
90°
270°
120°
240°
150°
210°
180°
tube

Fig2

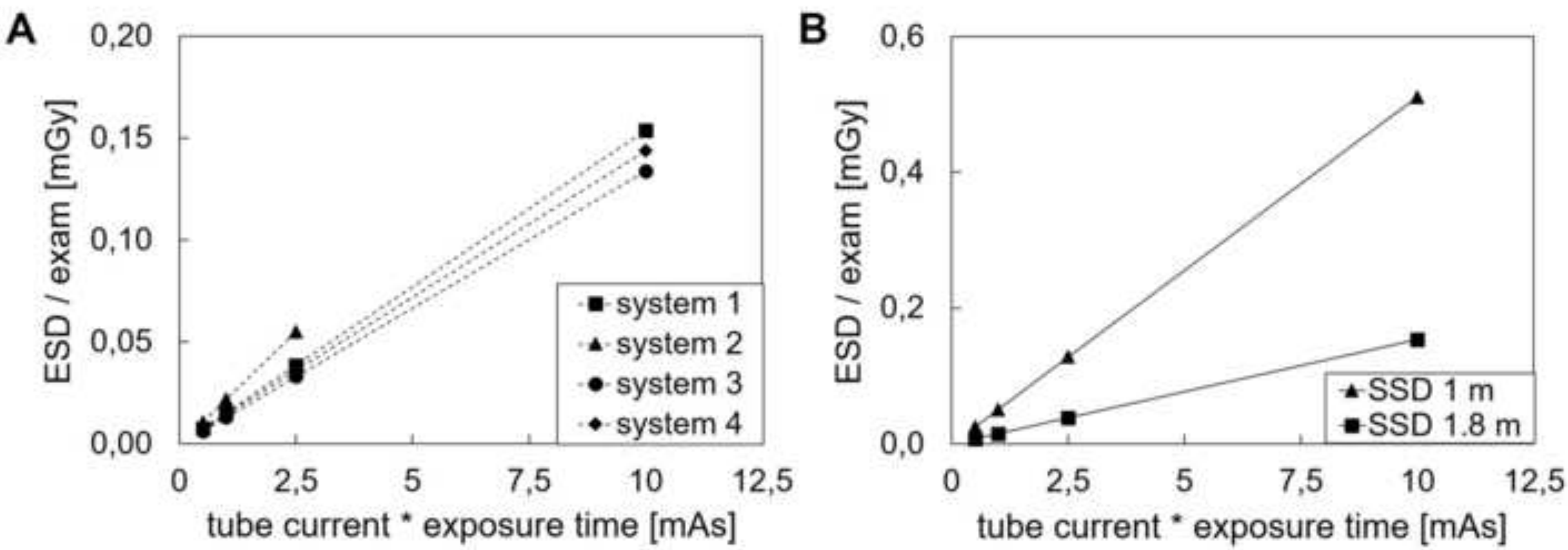
A
0,20
0,15
0,10
0,05
0,00
ESD / exam [mGy]
0
2,5
5
7,5
10
12,5
tube current * exposure time [mAs]
system 1
system 2
system 3
system 4
B
0,6
0,4
0,2
0,0
ESD / exam [mGy]
0
2,5
5
7,5
10
12,5
tube current * exposure time [mAs]
SSD 1 m
SSD 1.8 m

Fig3

Click here to access/download;Figure;Fig3_FIN.tiff

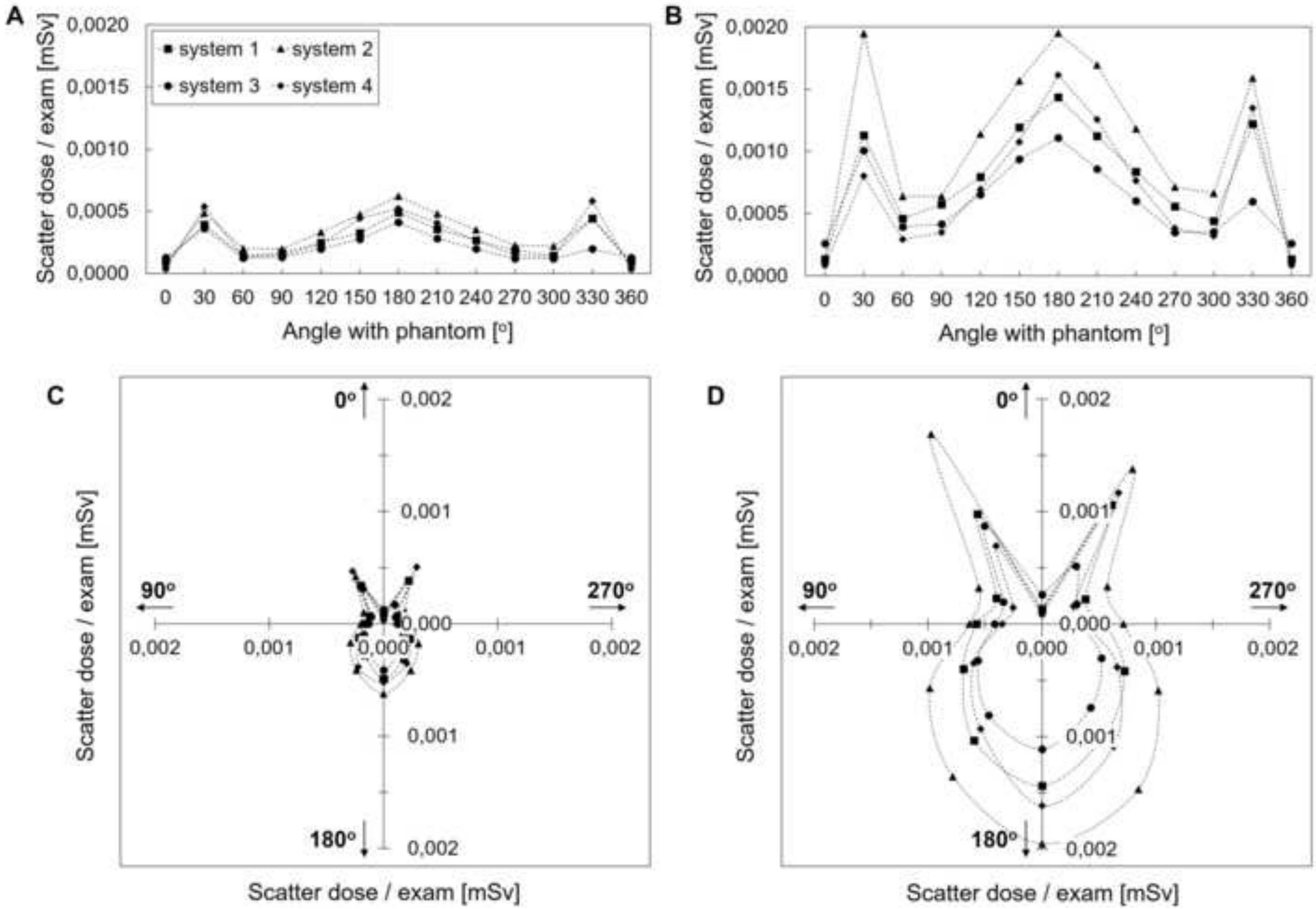

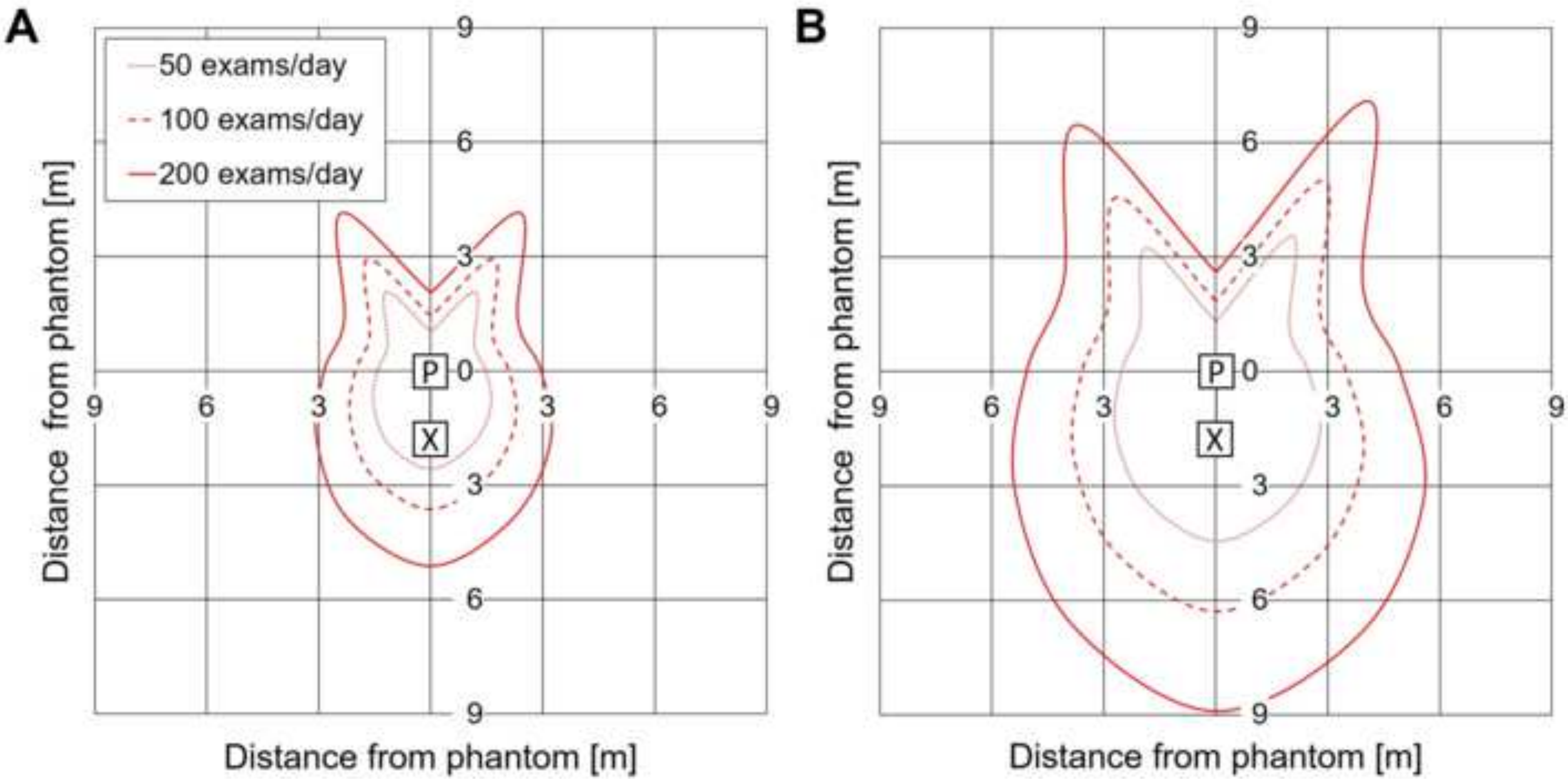
A
B
50 exams/day
100 exams/day
200 exams/day
9
6
3
0
3
6
9
P
X
Distance from phantom [m]
Distance from phantom [m]

Fig5

Click here to access/download;Figure;Fig5_FIN.tiff

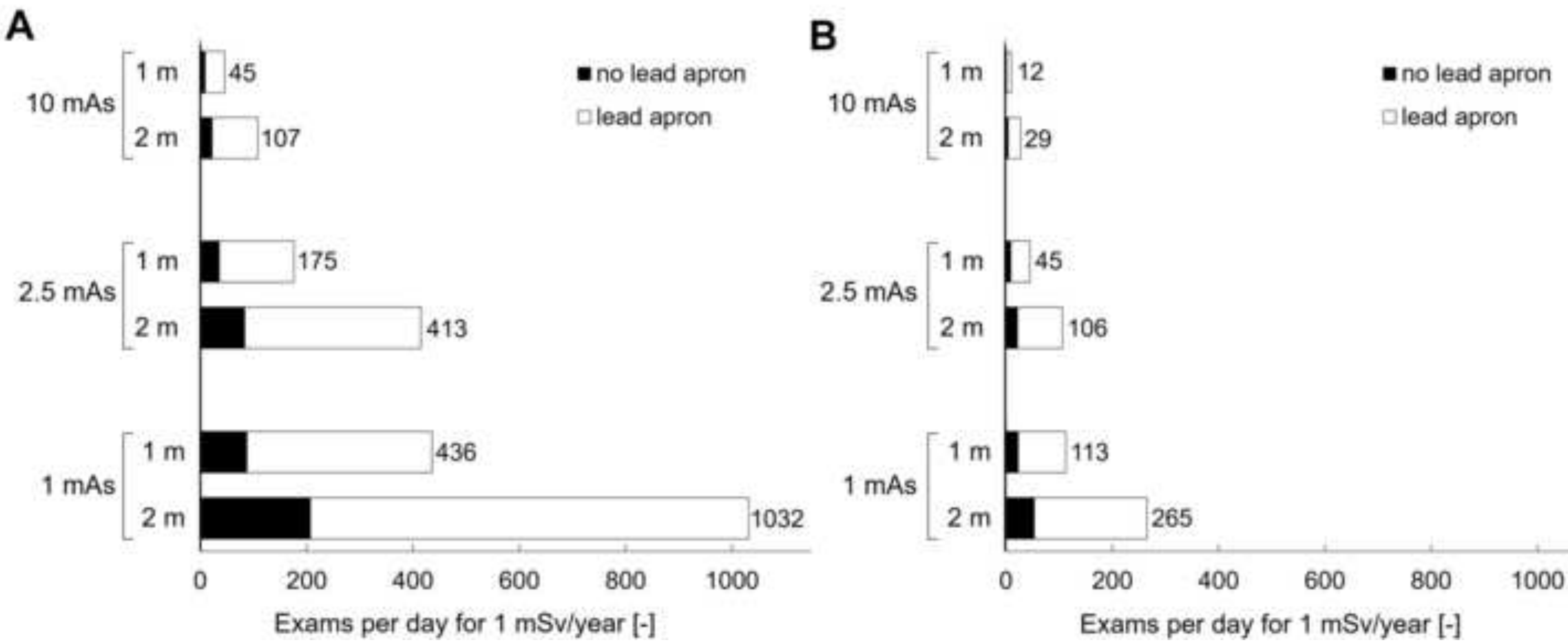

S1 Table

Click here to access/download
**Supporting Information**
S1 Table.pdf

S2 Table

Click here to access/download
**Supporting Information**
S2 Table.pdf

S3 Table

Click here to access/download
**Supporting Information**
S3 Table.pdf

Data Review URL

Click here to download Data Review URL

https://osf.io/4n8tk/?view_only=7db23094a11e46c09d0151b6f2e19a7c